\documentclass[%
 reprint,
superscriptaddress,
 amsmath,amssymb,
 aps,
longbibliography
]{revtex4-1}

\usepackage[colorlinks,urlcolor=blue,citecolor=blue,linkcolor=blue]{hyperref}
\usepackage{graphicx}
\usepackage{dcolumn}
\usepackage{bm}
\usepackage{hyperref}
\usepackage[mathlines]{lineno}
\usepackage[dvipsnames]{xcolor}

\usepackage{xspace}

\newcommand{\xiset}{\{\xi\}}
\newcommand{\IMRPhenomD}{{\texttt{IMRPhenomD}}\xspace}
\newcommand{\IMRPhenomC}{{\texttt{IMRPhenomC}}\xspace}
\newcommand{\SEOBNR}{{\texttt{SEOBNRv4\_ROM}}\xspace}

\newcommand{\apjs}{The Astrophysical Journal Supplement Series}
\newcommand{\cqg}{Classical and Quantum Gravity}
\newcommand{\pasp}{Publications of the Astronomical Society of the Pacific}

\newcommand{\Cardiff}{School of Physics and Astronomy, Cardiff University, Queens Buildings, Cardiff, CF24 3AA, United Kingdom}
\newcommand{\AEIHannover}{Max Planck  Institute for Gravitational Physics
(Albert Einstein Institute), Callinstr.~38, 30167 Hannover, Germany}
\newcommand{\UniHannover}{Leibniz Universit\"at Hannover, D-30167 Hannover, Germany}

\begin{document}

\preprint{APS/123-QED}

\title{Multi-waveform inference of gravitational waves}

\author{Gregory Ashton}
\affiliation{School of Physics and Astronomy, Monash University, Vic 3800, Australia}
\affiliation{OzGrav: The ARC Centre of Excellence for Gravitational Wave Discovery, Clayton VIC 3800, Australia}
\email{gregory.ashton@ligo.org}

\author{Sebastian Khan}
\affiliation{\Cardiff}
\email{sebastian.khan@ligo.org}
\affiliation{\AEIHannover}
\affiliation{\UniHannover}


\begin{abstract}

Bayesian inference of gravitational wave signals is subject to systematic error due to modelling uncertainty in waveform signal models, coined approximants. A growing collection of approximants are available which use different approaches and make different assumptions to ease the process of model development. We provide a method to  marginalize over the uncertainty in a set of waveform approximants by constructing a mixture-model multi-waveform likelihood. This method fits into existing workflows by determining the mixture parameters from the per-waveform evidences, enabling the production of marginalized combined sample sets from independent runs.
\end{abstract}

\maketitle

\section{\label{sec:introduction}Introduction}
Numerical relativity simulations of binary black hole mergers solve the full Einstein equations numerically and thus provide the most accurate predictions for the gravitational wave signal from compact binary coalescence (CBC) events \cite{PhysRevLett.95.121101,PhysRevLett.96.111101,PhysRevLett.96.111102,PhysRevD.77.024027}. These simulations are computationally demanding
and the requirement by stochastic parameter estimation methods
(see, e.g., \citep{veitch2015, lalsuite, 2019PASP..131b4503B, bilby2019, pankow2015, lange2018})
to rapidly generate the waveform at an arbitrary point within the prior space makes their direct use impractical, except in grid-based methods~\citep{pankow2015, lange2018, PhysRevD.94.064035} where the simulations can be pre-computed.
To remedy this, a growing collection of rapidly computable \emph{waveform approximants} for CBC signals have been developed, \cite{PhysRevD.59.084006,Pan:2013rra,PhysRevD.62.064015,Pan:2011gk,Taracchini:2013rva,Blanchet2014,Hannam:2013oca,Ajith:2007qp,PhysRevD.82.064016,Khan:2015jqa,PhysRevLett.120.161102,PhysRevD.98.084028,PhysRevD.95.044028,PhysRevD.95.024010,PhysRevD.96.024058, PhysRevD.98.104052,Mehta:2019wxm, Varma:2018mmi, Williams:2019vub,Setyawati:2018tqp,Khan:2018fmp,2019arXiv191106050K}, some of which are tuned to the numerical relativity simulations.

Typically, inference workflows proceed by first identifying a set of waveforms relevant to the expected signal based on the signal characteristics identified by the search pipelines (see Ref.~\citep{gwtc1} for an overview of the search process). Then, inference is run for each waveform resulting in a set of \emph{posterior samples}~\citep{veitch2015}. Differences between the inferred posteriors for each waveforms are understood to be due to the systematic differences in the waveform approximants; to create a set of results which are robust to these systematic waveform uncertainty, the naive-mixing method (used in, e.g., \citep{gw150914_properties, gwtc1}) is to combine equal numbers of samples from the posterior of each waveform into a single combined data set~\citep{QuotingPE}.

The choice to combine equal numbers of samples from multiple waveforms constitutes an equal-weighted probability on the waveform aproximants. In the absence of additional information, this may appear to be the only choice. However, there exists additional information in the quality of the waveform fit to the data: intuitively the idea presented in this work is to weight the samples by the computed posterior evidence. In Sec.~\ref{sec:method}, by treating the set of approximants as a mixture model, we show how the fit of the waveforms themselves to the data can be used to infer the appropriate mixing fraction and combine samples. In Sec.~\ref{sec:uncertain}, we discuss the effect of uncertainty on the evidence estimates and in Sec.~\ref{sec:toy-model} we provide a toy-model to help build intuition. We demonstrate that this method reduces waveform uncertainty by running an injection and recovery simulation in Sec.~\ref{sec:example} and apply the method to GW150914~\citep{gw150914_detection} in Sec.~\ref{sec:application}. We conclude with a discussion in Sec.~\ref{sec:discussion}.

\section{\label{sec:method}Method}
Given a set of $N$ waveforms $\{w_\ell\}$ with equivalently defined model parameters $\theta$, our goal is to compute $P(\theta| d, \{w_\ell\})$, the posterior distribution conditional on both the data $d$ and the set of waveforms. First, let us associate to each waveform a hypothesis $H_\ell$ that the data was generated with the $\ell$th waveform; the hypothesis includes prior-choices for the model parameters $\theta$. 

To obtain the likelihood for some data $d$, we assume that the hypotheses are exhaustive such that $1 = \sum_{\ell} P(H_\ell| d, \theta, \xiset)$ where $\xiset$ are a set of prior-probability hyperparameters for each hypothesis, $\xi_\ell \equiv P(H_\ell | \xiset)$. For unitarity, we require $\xi_N = 1 - \sum_{\ell=0}^{N-1} \xi_\ell$. The likelihood can now be written as a mixture model with mixing parameters $\xi_\ell$:
\begin{align}
    P(d| \theta, \xiset) = \sum_{\ell} P(d | H_\ell, \theta) \xi_\ell\,.
    \label{eqn:likelihood}
\end{align}
This multi-waveform likelihood can be used in place of the usual likelihood (see, e.g. \citet{veitch2015}) to perform multi-waveform inference. (Note that, if used in practise, computing the likelihood for each waveform serially will slow down the per-likelihood compute time; the computation of $P(d| H_\ell, \theta)$ should instead be parallelised to reduce the overall compute time).

While the multi-waveform likelihood is simple to implement, it does not fit into the typical existing workflows described above. The solution is to first run inference for each waveform independently, then infer the posterior mixing fraction
\begin{align}
    \hat{\xi}_{\ell} \equiv P(H_\ell | d) = \int d\theta P(H_\ell, \theta | d)\,,
    \label{eqn:xihatl}
\end{align}
where we use a ``hat'' to distinguish $\hat{\xi}_\ell$ as the posterior mixing fraction for the $H_\ell$ model. The set $\{\hat{\xi}\}$ sum to unity and can be used to determine the mixing weights which should be applied to posterior samples $P(\theta | H_\ell, d)$. If an equal-weighted prior probability is assigned to each waveform, then
\begin{align}
     \hat{\xi}_{\ell} =  \frac{Z_\ell}{\sum_j Z_{j}}\,,
     \label{eqn:xihatl-calculation}
\end{align}
where
\begin{align}
    Z_\ell \equiv P(d| H_\ell) = \int d\theta\, P(d| H_\ell, \theta) P(\theta| H_\ell)\,,
    \label{eqn:evidence}
\end{align}
is the per-waveform evidence.

For each individual waveform, we run inference and produce posterior samples. Then, instead of combining the samples equally, we combine them with weights given by Eq.~\eqref{eqn:xihatl-calculation}. This yields a combined set of samples appropriately marginalized over the set of input waveforms. It is worth stating that this is not the same as marginalizing over waveform uncertainty in general: only uncertainty conditional on the input set of waveforms is captured.

This method of combining samples can be used when different waveform-hypotheses imply different priors on the model parameters. This can be seen in Eq.~\eqref{eqn:evidence}, where the model-parameter prior, $P(\theta | H_\ell)$, is conditional on the waveform-hypothesis $H_\ell$. Because of this feature, samples from seemingly different waveform types can be combined, provided they refer to the same set of underlying model parameters, but with a differing prior. For example, if $w_1$ is a waveform including the tidal deformability parameters, $\lambda_1, \lambda_2$, this can be combined with samples from $w_2$, if $w_1$ and $w_2$ are equivalent when the tidal deformability parameters tend to zero. In this case, the prior on the tidal deformability parameters for $w_2$ are Dirac delta functions with peaks at zero.

\section{Uncertain mixing fractions}
\label{sec:uncertain}

The method outlined in Sec.~\ref{sec:method} assumes that the mixing fractions can be calculated exactly. In practise, for gravitational wave signals, we do not have a closed form expression for Eq.~\eqref{eqn:evidence}. Instead, we estimate the evidence and posterior distribution using stochastic sampling methods \citep{veitch2015, lalsuite, 2019PASP..131b4503B, bilby2019, pankow2015, lange2018}. These typically yield an estimated log-evidence with some uncertainty (usually expressed as an uncertainty on the log-evidence). For a discussion on how this is derived for the \texttt{dynesty} sampler used in this work, see \citet{dynesty}. 

If the log-uncertainty is sufficiently small with respect to the differences between log-evidences, it can of course be neglected. However, in cases where this is not true, care must be taken: ignoring the log-evidence uncertainty will result in an overly constrained posterior.

To include the uncertainty on the log-evidence into the mixing process, first we must define the distribution of evidences. For the \texttt{dynesty} sampler, we show in Appendix~\ref{app:evidence} that $\log({Z})\sim \mathrm{Normal}(\log Z', \sigma_{\log (Z)})$, i.e. the log-evidence is normally distributed with mean given by the estimated log-evidence $\log Z'$ and standard deviation given by the estimated uncertainty $\sigma_{\log(Z)}$.

With an appropriate parametric form of the distribution of evidences, we then combine samples by the following process. (\emph{a}) Draw a set of evidences for each waveform from their estimated distribution, (\emph{b}) calculate the weights from Eq.~\eqref{eqn:xihatl-calculation}, (\emph{c}) apply the weights when drawing a sample from the combined set of samples. This process is then repeated until a sufficient number of samples are drawn for the mixed posterior. Removing step (\emph{a}) of course reduces the operation to that defined in Sec.~\ref{sec:method} and is appropriate if the uncertainty on the evidences is sufficiently small.

\begin{figure}
    \centering
    \includegraphics[width=0.45\textwidth]{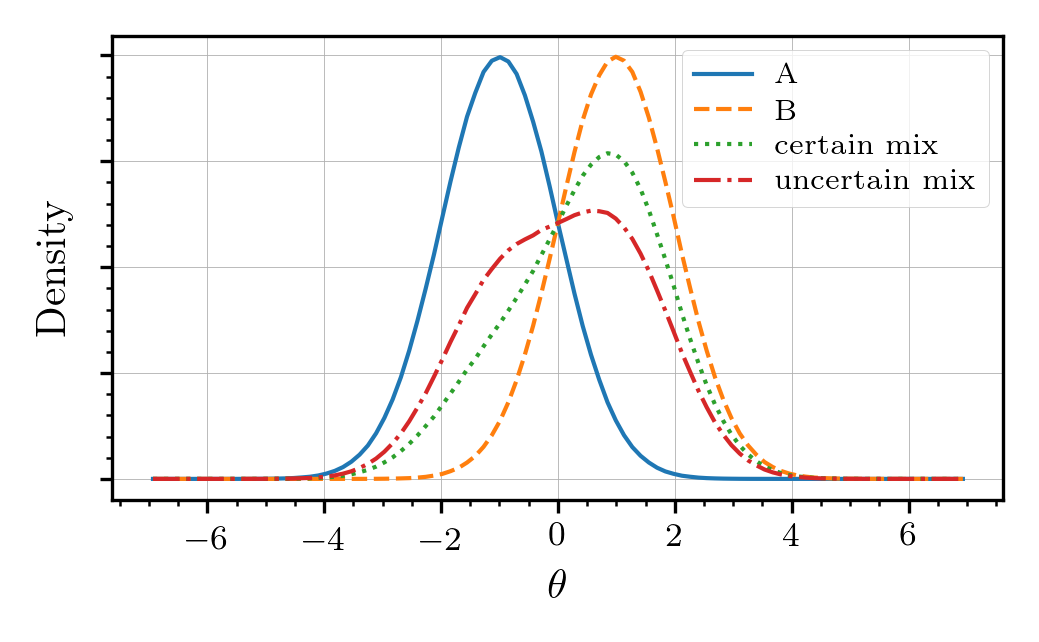}
    \caption{A comparison of the mixed posterior for an arbitrary parameter $\theta$ from two distributions when neglecting the evidence uncertainty, ``certain mix'', and when including it, ``uncertain mix'', using the procedure outlines in Sec.~\ref{sec:uncertain}.
    }
    \label{fig:uncertain-mixing-fraction}
\end{figure}

What effect does uncertainty in the evidence calculations have on the mixed posterior? To investigate this question, we perform a simple numerical experiment: we draw two sets of posterior samples from $P(\theta| A) \sim \mathrm{Normal(-1, 1)}$ and $P(\theta| B) \sim \mathrm{Normal(1, 1)}$ where $\theta$ is an arbitrary unknown variable which we wish to estimate and $A$ and $B$ arbitrarily label the posterior from two ``waveforms''. We then defined that $\log(Z_{A}/Z_{B}) = 1$, but that each the log-evidences themselves have a normally distributed uncertainty with standard deviation $\sigma_{\log(Z)}=5$. Following the procedure outlined above, we mix the posteriors. The results are given in Fig.~\ref{fig:uncertain-mixing-fraction} for both an ``uncertain mix'' where we include the uncertainty on the evidences and ``certain mix'' where we neglect that uncertainty.

Fig.~\ref{fig:uncertain-mixing-fraction} illustrates that including the additional uncertainty on the evidences produces a more conservative estimate. The uncertain mix posterior is less skewed to the $B$ posterior than the certain mix. We note that repeating the numerical experiment, but in a case where $\log(Z_{A}/Z_{B})=0$, i.e. the Bayes factor between the two posteriors is equally weighted, we find that the uncertain mixture and the certain mixture are indistinguishable for any amount of uncertainty in the evidences. 

This numerical study suggests it is prudent to include the effects of uncertainty in the evidence estimates using the procedure outlined above. This adds little computational complexity, provided it is easy to sample from the distribution of evidences.

\section{Toy model}
\label{sec:toy-model}
To build intuition about the method, we describe here a simple toy model consisting of a sinusoidal function with a linearly-varying angular frequency
\begin{align}
    y(t) = \sin\left(\omega t + \frac{1}{2}\dot{\omega} t^2\right)\,.
\end{align}
We then define two ``waveforms'' consisting of a choice for the rate of change of angular frequency:
\begin{align}
    w_A & \rightarrow \dot{\omega}=0 \\
    w_B & \rightarrow \dot{\omega}=0.1
\end{align}

\begin{figure}[tb]
    \centering
    \includegraphics{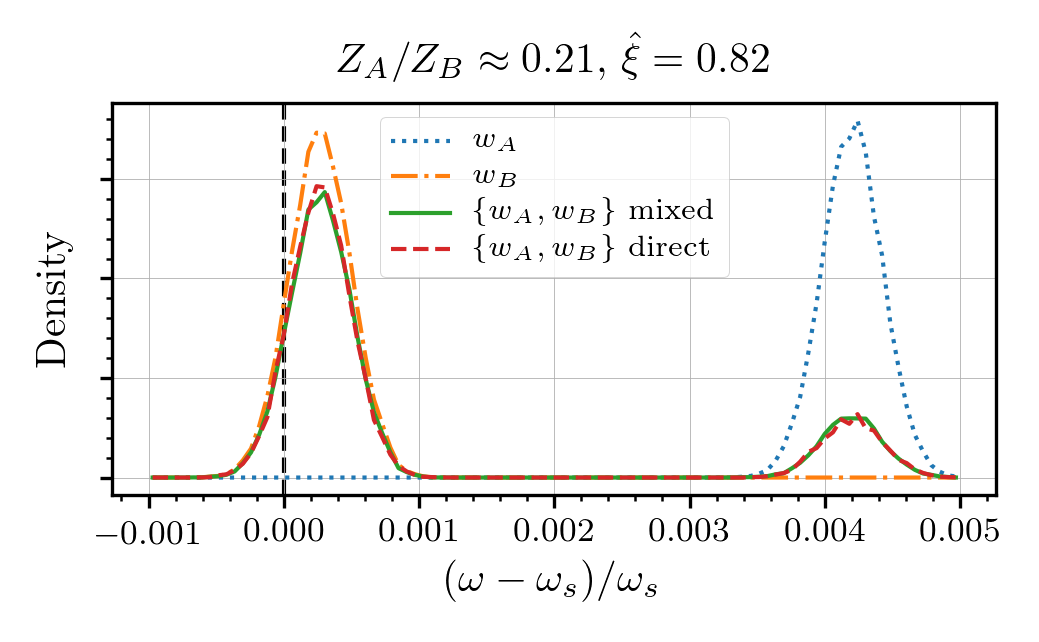}
    \caption{Posterior distribution on the angular frequency $\omega$ as a relative offset compared to the simulated value $\omega_s$ for each waveform separately, and the set of both waveforms. The term ``mixed'' refers to calculating the posterior mixing fraction using Eq.~\eqref{eqn:xihatl-calculation}; we apply the procedure in Sec.~\ref{sec:uncertain} to include the numerical uncertainty on the estimated evidence. The term  ``direct'' refers to applying Eq.~\eqref{eqn:likelihood} directly. In the title, we give the Bayes factor between the models and the numerical value of $\hat{\xi}$ neglecting the uncertainty.} 
    \label{fig:toy-model}
\end{figure}

We simulate data consisting of waveform $w_B$ with Gaussian noise of known standard deviation $\sigma=0.01$. We then then apply inference for $w_A$ and $w_B$ separately, producing two disjoint posteriors (see Fig.~\ref{fig:toy-model}). The goal of this work is to present a method for combining samples between waveforms. Given the binary choice between the two toy-model waveforms, we have two options to do this. We can mix the posteriors using the posterior mixing fractions, Eq.~\eqref{eqn:xihatl-calculation}, or we can apply the likelihood of Eq.~\eqref{eqn:likelihood} directly and perform inference on the mixture model itself.

For the simulated data with $\sigma=0.01$, the Bayes factor between the two waveforms is $Z_B/Z_A\approx5$: indicating a preference for $w_B$, but not overwhelmingly so. 
In Fig.~\ref{fig:toy-model}, we show the posterior distribution on the only unknown model parameter $\omega$ for four cases (see caption).
That the inference when using only $w_A$ is biased is expected since the data was simulated using $w_B$. This case was specifically chosen to illustrate the behaviour when the data is not sufficiently informative to rule out the erroneous waveform, $\omega_B$.

The results mixing the posteriors according to Eq.~\eqref{eqn:xihatl-calculation} and applying the likelihood Eq.~\eqref{eqn:likelihood}, to within the sampling errors, demonstrate equivalent posteriors. This validates that the mixing process is equivalent to direct inference.

The combined posterior in Fig.~\ref{fig:toy-model} (from either the mixing or direct methods) is multi-modal. This is a proper reflection of the posterior uncertainty on the model parameter: each mode is inherently associated with a different model. This is happening in this special case because the evidences are not especially informative: the Bayes factor only demonstrates a mild preference for $\omega_B$.

\begin{figure}[tb]
    \centering
    \includegraphics{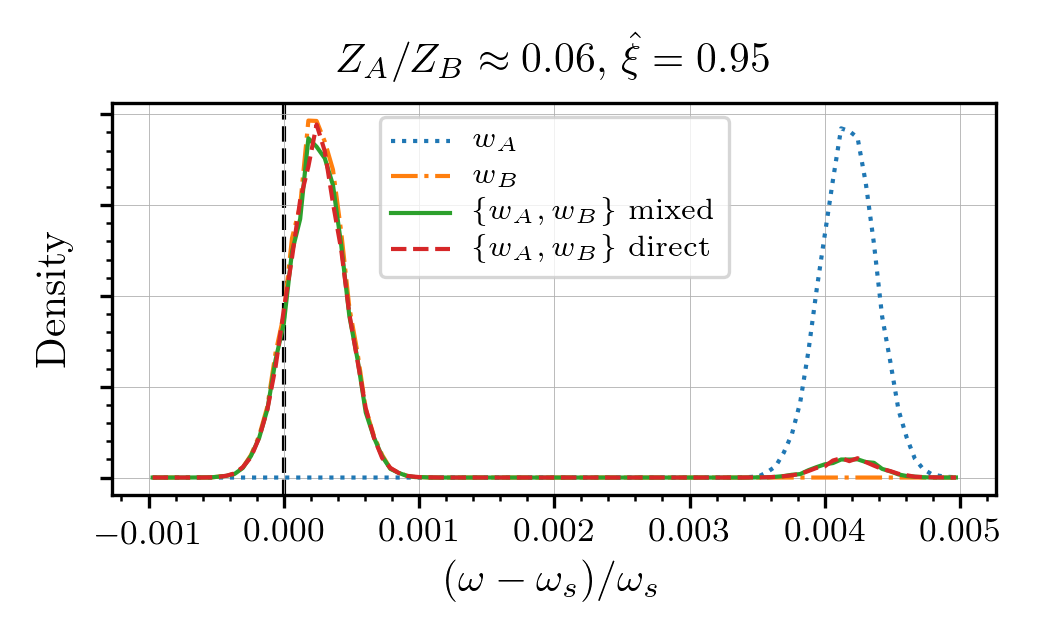}
    \caption{Repeating the toy simulation of Fig.~\ref{fig:toy-model}  (see caption for details), but with a smaller level of noise in the data: the reduction in noise leads to more informative evidence estimates and demonstrates the mixed posterior converging to the true value while neglecting the erroneous model A.} 
    \label{fig:toy-model-high-snr}
\end{figure}

To gain intuition when the evidences become more information, in Fig.~\ref{fig:toy-model-high-snr}, we repeat the toy model of Fig.~\ref{fig:toy-model}, but decreasing the level of noise. In particular, we reduce the standard deviation $\sigma$ from 0.01 to 0.009. With this modest decrease in the noise the Bayes factor is now four times more in favour of $B$ than was the case for Fig.~\ref{fig:toy-model}. Correspondingly, the mixing fraction increases and the posterior is now more weighted to the correct $B$ mode. Further decreasing the noise, the $A$ mode is eventually (for Bayes factor above $\sim 50$ or so) entirely ruled out. On the other hand, if we were to increase the level of noise, we would see a more equal mixing between the two since the evidence would favour neither one or the other.

\section{Injection and recovery}
\label{sec:example}
To demonstrate the utility of this method, we now run a simple injection and recovery test. Aligned-spin signals generated by the \IMRPhenomD~\citep{PhysRevD.93.044006,Khan:2015jqa}  waveform model are added to simulated coloured-Gaussian data from two detectors (Hanford and Livingston) with Advanced LIGO design sensitivity \citep{aLIGO_sensitivity, LIGO}. The data is simulated and analysed using the \texttt{Bilby} \citep{bilby2019} Bayesian inference software. The simulated source parameters are generated by random draws from the prior (see Table~\ref{tab:parameters}); repeated draws are made until the network optimal signal to noise ratio (SNR) exceeds a threshold of 8, a typical search threshold.

\begin{table}[tb]
    \centering
    \begin{tabular}{lrlll}
    \hline \hline
         Parameter & \multicolumn{3}{c}{Prior support} & \\\hline
         Chirp mass $\mathcal{M}$& 25 &--& 100 & $M_{\odot}$\\
         Mass ratio $q$& 0.125 &--& 1 & \\
         Primary spin $\chi_{1, z}$& -0.9 &--& 0.9 & \\
         Secondary spin $\chi_{2, z}$& -0.9 &--& 0.9 & \\
         Lum. distance $d_l$& 0.1 &--& 5 & Gpc\\
         Inclination $\theta_\mathrm{JN}$ & 0 &--& $\pi$ & rad.\\
         Right Asc. $\alpha$ & 0&--& $2\pi$ & rad.\\
         Declination $\delta$ & -$\pi/2$ &--& $\pi/2$ & rad.\\
         Polarisation angle $\psi$ &0 &--& $\pi$ & rad \\
         Phase $\phi$ & 0 &--& $2\pi$ & rad.\\
         Geocentric-time $t_c$ & -0.1 &--& 0.1 & s\\ \hline \hline
    \end{tabular}
    \caption{Prior support for the source parameters. $\chi_{1, z}$ and $\chi_{2,z}$ use the ``z-prior'' (see Eq.~(A7) of \citet{lange2018}); the luminosity distance prior is $P(d_L) \propto d_L^2$; the inclination angle is Cosine distributed; the declination is Sine distributed; all other parameters are Uniformly distributed. The geocentric time, $t_c$ is given relative to the simulated trigger time.}
    \label{tab:parameters}
\end{table}

\begin{figure}
    \centering
    \includegraphics[width=0.45\textwidth]{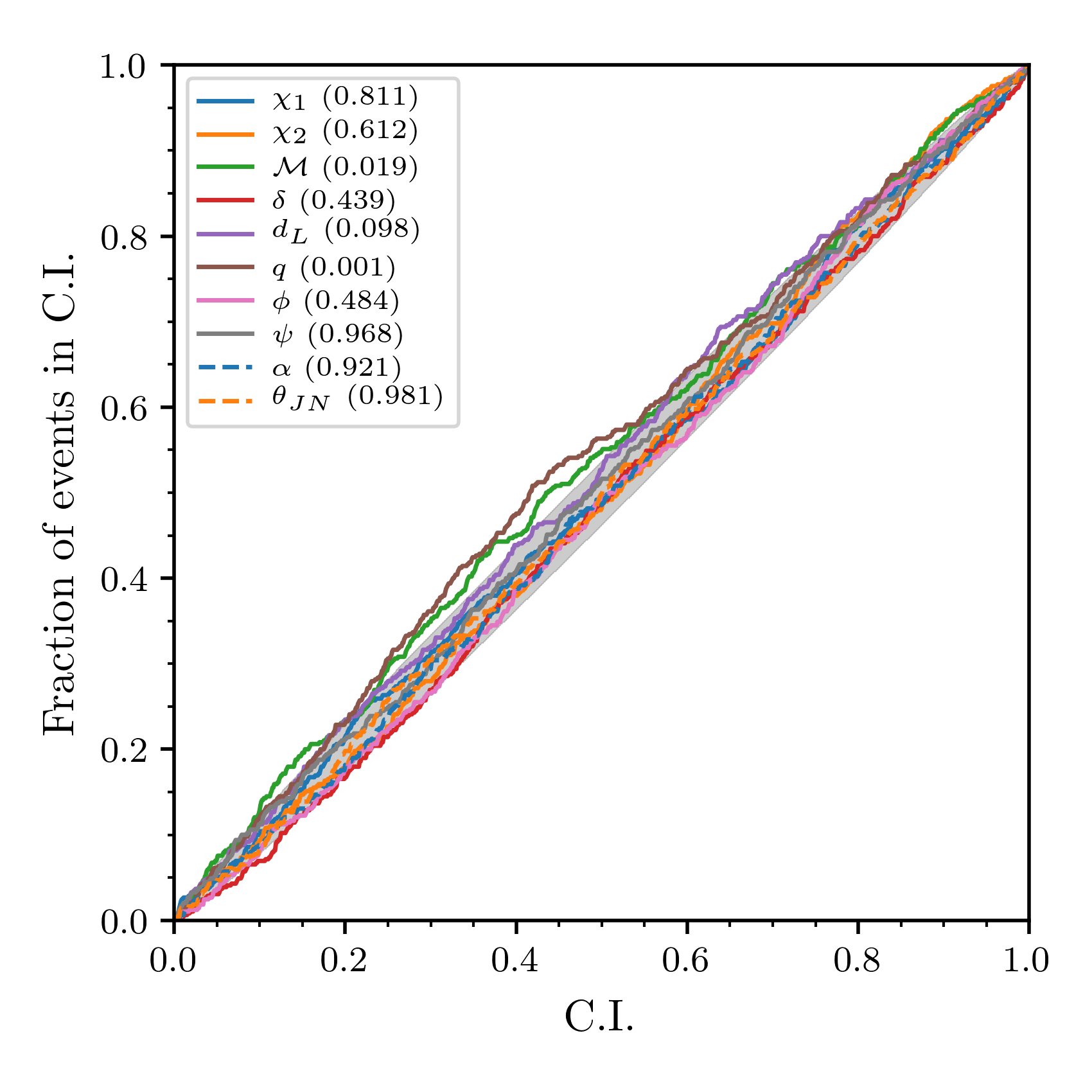}
    \caption{We simulate 500 \IMRPhenomD signals in advanced-LIGO design sensitivity noise \citep{aLIGO_sensitivity}. Parameter estimation is performed on all simulated data sets using both the \IMRPhenomD and \IMRPhenomC waveforms and the resulting samples are mixed using the informed-mixture method, i.e. with mixture parameters calculated from Eq.~\eqref{eqn:xihatl-calculation}. This figure shown the pp-test diagnostic plot applied to the 500 mixed posteriors.
    }
    \label{fig:pp-test}
\end{figure}

For each simulated signal, we recover with either \IMRPhenomC~\citep{PhysRevD.82.064016} or \IMRPhenomD. To probe the inherent bias, we repeat this process on 500 simulated data sets and perform a percentile-percentile (pp-test) (based on the work of \citet{cook2006}). Graphically a pp-test is a plot of the fraction of signals with true parameter recovered to within a credible interval against the credible interval itself (we show an example later in Fig.~\ref{fig:pp-test}). The pp-test is a useful diagnostic for investigating bias: a pass in a pp-test verifies that the posterior recovery is unbiased with respect to the injections (i.e. the $x$\% posterior intervals contain the true values $x$\% of the time).
For any pp-test, a summary statistic is obtained by first calculating a $p$-value from the Kolmogorov-Smirnov test on each parameter separately, then combining these together using Fisher's method. The resulting set of $p$-values (for which the null hypothesis is that the results are unbiased) are given in Table~\ref{tab:p-values}. 

\begin{table}[ht]
    \centering
    \begin{tabular}{clr}
    \hline \hline
        Injection & Recovery & $p$-value \\\hline
        \IMRPhenomD & \IMRPhenomD & 0.45 \\
        \IMRPhenomD & \IMRPhenomC & $< 0.01$ \\
        \IMRPhenomD & informed-mix & $0.045$ \\
        \IMRPhenomD & naive-mix & $0.0092$ \\ \hline \hline
    \end{tabular}
    \caption{Table of combined p-values calculated for each injection-recovery test case. Combinations are made over the full set of parameters in Table~\ref{tab:parameters}, except the geocentric time which has known systematic shifts between waveforms.}
    \label{tab:p-values}
\end{table}

For the case when the injection and recovery are performed using the same waveform, as expected we find a $p$-value indicating the posteriors are unbiased: this demonstrates that in the absence of systematic differences in the injections and recovery waveforms, the underlying method (i.e. the generation of injection values and posterior sampling) is unbiased.

On the other hand, when the recovery waveform (\IMRPhenomC) differs from the injection, the $p$-value is small, indicating the results are biased.  Per-parameter analysis indicates that it is the merger time, mass ratio, and chirp mass parameters which fail the test. The cause for this, systematic differences between waveforms, is well understood and expected~\citep{orca93841}.

The true signal in the simulated data set is \IMRPhenomD. However, we now consider the case when we have uncertainty about which waveform best approximates the signal. Using the method described in Sec.~\ref{sec:method}, the set of posterior samples conditional on both waveforms is obtained by mixing together samples from the \IMRPhenomC- and \IMRPhenomD-recovery with a mixing fraction given by the ratio of their evidence to the total evidence, Eq.~\eqref{eqn:xihatl-calculation}. Repeating this for each simulated data segment, we apply the pp-test to the resulting samples, the pp-test plot itself is given in Fig.~\ref{fig:pp-test} and the combined $p$-value is labelled as ``informed-mix'' in Table~\ref{tab:p-values}. The results are biased, but to a substantially lesser extent than for \IMRPhenomC alone. 

That the $p$-value indicates a bias for the informed-mixture biased is unsurprising since the pp-test is only expected to pass when the data-generation exactly matches the assumptions of the model-fitting software \citep{cook2006}; in this case we have introduced additional uncertainty into the model-fitting.

Mixing the samples according to Eq.~\eqref{eqn:xihatl-calculation} is the better thing to do, given uncertainty about the waveform. The Gravitational Wave Transient Catalogue \citep{gwtc1}, used the naive-mixing method, combining equal numbers of posterior samples from multiple waveforms. We implement this ``naive-mix'' method and apply it to the set of samples from each individual waveform. The resulting $p$-value is smaller than the informed-mix, indicative of a greater degree of bias. Nevertheless, it demonstrates that for cases where the waveforms make highly similar predictions (a statement that can be quantified by a Bayes factor between them), the naive-mixing method is reasonable, but should only be used when the evidence calculations are infeasible.

We have investigated here the typical case for advanced-era CBC detections in which both waveforms perform reasonably well in fitting the data (see the next Section for a demonstration). Had the set of injections (or sensitivity of the simulated instruments) been such that the differences in waveforms were more apparent, the informed mixing fraction would preference the injected waveforms and mix the posterior samples accordingly. As an example, consider the case where two waveforms ($w_A$ and $w_B$) are applied and each produces a set of $10^4$ posterior samples. For the probability of including any samples from $w_B$ to be less than 1, $\hat{\xi}<10^{-4}$, which implies the Bayes factor between them must be $Z_A/Z_B \lesssim 10^{-4}$. For such a case, the posterior samples would (almost) all be drawn from $w_A$. If repeated in a pp-test, the informed-mixture method would be unbiased, but the naive-mixture method would not. 

\section{Application to GW150914}
\label{sec:application}
To apply the method in practise, we run Bayesian inference on the first-observed binary black hole coalescence, GW150914~\citep{gw150914_detection,gw150914_properties, LIGO}. This system has been well studied and posterior samples are available \citep{gw150914_properties, gwtc1}, but the evidences are not. These original analyses used both precessing and non-precessing waveform approximants. Here, as an illustrative example of the effect of waveform-approximant mixing only, we perform analysis for three non-precessing waveform approximants, \IMRPhenomC \citep{PhysRevD.82.064016}, \IMRPhenomD~\citep{PhysRevD.93.044006,Khan:2015jqa}, and \SEOBNR~\citep{PhysRevD.95.044028}.
The analysis is done using \texttt{Bilby} \citep{bilby2019} on data from the Gravitational Wave Open Science Centre \citep{gwosc}\footnote{Gravitational Wave Open Science Center (GWOSC), \url{www.gw-openscience.org/events/GW150914/}, \href{https://doi.org/10.7935/K5MW2F23}{DOI:10.7935/K5MW2F23}} following the methodology described in Appendix B of Ref.~\citep{gwtc1}. 

The evidences computed for each waveform can be used to construct Bayes factors
\begin{align}
    \ln\left(\frac{Z_\mathrm{\IMRPhenomD}}{Z_\mathrm{\SEOBNR}}\right) = 0.02\pm 0.5\,,
    \label{eqn:bfA}
\end{align}
and 
\begin{align}
    \ln\left(\frac{Z_\mathrm{\IMRPhenomD}}{Z_\mathrm{\IMRPhenomC}}\right) = 0.5\pm 0.5\,.
    \label{eqn:bfB}
\end{align}
These results confirm what is known in the literature (see, e.g., \citep{gw150914_properties,payne2019}): GW150914 and other events seen in the first and second observing runs of LIGO and Virgo are not sufficiently loud to decisively distinguish waveform approximants. 

\begin{figure}[htb]
    \centering
    \includegraphics[width=0.5\textwidth]{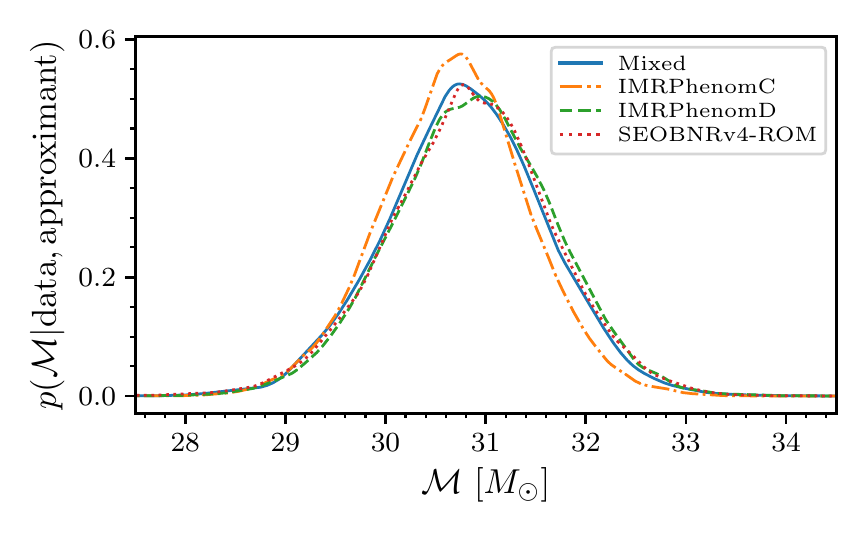}
    \caption{fThe posterior probability density for the detector-frame chirp mass, $\mathcal{M}$, in solar masses for three aligned-spin waveform approximants and their mixture, applying Eq.~\eqref{eqn:xihatl-calculation}and the procedure for including uncertainty in the evidences from Sec.~\ref{sec:uncertain}. The same uniform prior on detector-frame chirp mass was applied for all three individual waveform approximants.}
    \label{fig:chirp}
\end{figure}

The mixing fractions for these three approximants, applying, Eq.~\eqref{eqn:xihatl-calculation}, are $0.24$, $0.39$, and $0.38$ for the \IMRPhenomC, \IMRPhenomD, and \SEOBNR waveforms respectively. To illustrate the effect of the mixing, in Fig.~\ref{fig:chirp}, we plot the posterior probability density for the detector-frame chirp mass from the three aligned-spin waveform approximants and the mixture. Of the three waveforms, \IMRPhenomC has the smaller evidence and only a quarter of the samples are drawn from this posterior: as a result, the mixture is closer to the \IMRPhenomD and \SEOBNR posteriors.

Because the difference in evidences between the waveforms is small, as demonstrated by the Bayes factors in Eq.~\eqref{eqn:bfA} and \eqref{eqn:bfB}, the informed-mixing method will yield results similar to those produced by the naive-mixing method. With future detections, when the data is more informative about features in the waveform approximants, we strongly recommend that the informed-mixing method is applied when combining samples to ensure the results properly reflect the posterior uncertainty.

\section{Conclusion and Outlook}\label{sec:discussion}
The informed-mixing method presented here provides an improvement on the naive-mixing method to combine samples from multiple waveforms by including information from the estimated evidences about how well the waveforms fit to the data. Ultimately, this should make it easier to include multiple waveforms
without concern about the introduction of biases due to sub-optimal combination of posterior samples.

To use this optimal method for multi-waveform inference, accurate estimation must be made of the waveform evidence, Eq.~\eqref{eqn:evidence}. As such, the ability to properly handle systematic uncertainty in the waveform is critically underpinned by the ability estimate the evidence. We encourage future analyses of CBC systems to ensure evidence estimates are calculated and reported.
In Sec.~\ref{sec:uncertain}, we outline a procedure to include the uncertainty on these evidence estimates into the mixing process.

The mixture-model method presented in this work was discussed in the context of multi-waveform inference. Another systematic uncertainty is in the estimate of the power spectral density (PSD) used to characterise the detector noise~\citep{Rover:2008yp}. The state of the art method (used in \citet{gwtc1}) involves applying the \texttt{BayesLine} algorithm \citep{Cornish:2014kda, Littenberg:2015} which computes a posterior probability distribution for possible PSDs, then using the median PSD in PE analyses~\citep{Chatziioannou:2019zvs}. However, marginalizing over the uncertainty, rather than making a point estimate of the PSD is preferable. The methodology presented in Sec.~\ref{sec:method} can be applied to this problem. If multiple runs are performed with differing draws from the \texttt{BayesLine} posterior, Eq.~\eqref{eqn:xihatl-calculation} can be applied to calculate mixing fractions with which to combine posteriors. This method of PSD marginalization is sub-optimal compared to the general method of fitting both the PSD and source model simultaneously \citep{Chatziioannou:2019zvs}, however it makes the problem of marginalizing over PSD uncertainty embarrassingly parallel. 

\begin{acknowledgments}
We thank Christopher Berry, John Veitch, Michael P{\"u}rrer, and members of the LIGO and Virgo Compact Binary Coalescence group for valuable input during the preparation of this manuscript. We also thank the anonymous referee for valuable comments which improved the manuscript during review. G.A.~is supported by the Australian Research Council through grants CE170100004,  FT150100281, and DP180103155. S.K.~acknowledges support by the
Max Planck Society's Independent Research Group Grant. This research has made use of data, software and/or web tools obtained from the Gravitational Wave Open Science Center (https://www.gw-openscience.org), a service of LIGO Laboratory, the LIGO Scientific Collaboration and the Virgo Collaboration. LIGO is funded by the U.S. National Science Foundation. Virgo is funded by the French Centre National de Recherche Scientifique (CNRS), the Italian Istituto Nazionale della Fisica Nucleare (INFN) and the Dutch Nikhef, with contributions by Polish and Hungarian institutes. 
The authors are grateful for computational resources provided by the LIGO Laboratory and supported by National Science Foundation Grants PHY-0757058 and PHY-0823459. All analyse performed with \texttt{Bilby} in this work makes use of the \texttt{dynesty} \citep{dynesty} nested-sampling package. The \texttt{scipy}~\citep{scipy} and \texttt{matplotlib}~\citep{matplotlib} packages are used for statistical computations and creating figures.

\end{acknowledgments}

\bibliography{bibliography}

\appendix
\section{The distribution of uncertainties from dynesty}
\label{app:evidence}

In this Appendix, we demonstrate that the distribution of log-evidence errors from the \texttt{dynesty} \citep{dynesty} nested sampling package are normally distributed and conservatively bounded by the estimated log-evidence uncertainty. We do this using a one-dimensional problem. We cannot ensure that the same behaviour will occur in higher-dimensional problems, that will require further testing. But, this investigation is designed to give some intuition in a low-dimensional case where repeated numerical calculations are not computationally prohibitive.

To test the sampler, consider the integral
\begin{align}
    Z = \int_{\mu_\textrm{min}}^{\mu_\textrm{max}}
    \frac{1}{\sqrt{2\pi}}
    \exp^{-\frac{\mu^2}{2}}
    \frac{1}{\mu_\textrm{max} - \mu_\textrm{min}}
    \,d\mu\,,
    \label{eqn:Ztest}
\end{align}
which is the evidence for a Gaussian likelihood with $x=0$, unknown mean $\mu$, but known variance $\sigma^2=1$, and a uniform prior on $[\mu_\textrm{min}, \mu_\textrm{max}]$. If the uniform prior is sufficiently wide with respect to the posterior support, the integral is simply the Gaussian integral and hence the evidence has the approximate closed-form solution
\begin{align}
    Z \approx \frac{1}{\mu_\textrm{max} - \mu_\textrm{min}}\,.
    \label{eqn:Zclosedform}
\end{align}
Alternatively, Eq.~\eqref{eqn:Ztest} can be estimated using stochastic sampling software.

We use the \texttt{dynesty} sampler to compute the evidence and repeat the calculation 500 times whilst varying the number of live-points used by the sampler (a quantity proportional to the expected precision of the evidence estimate). In Tab.~\ref{tab:evidence}, we report summary statistics for the distribution of
\begin{align}
    f = \log{Z} + \log(\mu_\textrm{max} - \mu_\textrm{min})\,,
    \label{eqn:f}
\end{align}
the difference between the computed log-evidence and the logarithm of Eq.~\eqref{eqn:Zclosedform}, the approximate closed-form evidence.

That the mean values of Table~\ref{tab:evidence} are small relative to the typical errors indicates that the evidences themselves are not substantially biased. That the first estimate of the error $\sigma_0$ is close to, but always greater than, the empirically measured value $\sigma(f)$ suggests that the initial reported uncertainty on the evidences are conservative upper estimates. Finally, that the normality test $p$-values are all greater than a standard threshold of $0.05$ indicates that the distribution of $f$, and hence of $\log{Z}$ itself is consistent with a normal distribution.

\begin{table}[b]
    \centering
    \begin{tabular}{cllll}
    \hline \hline
        live-points & $\mu(f)$ & $\sigma(f)$ & $\sigma_0$ & $p$ \\\hline
        100 & 0.0032 & 0.095 & 0.111 & 0.97 \\
        250 & 0.0045 & 0.061 & 0.073 & 0.14 \\
        500 & 0.00032 & 0.047 & 0.053 &  0.38\\ \hline \hline
    \end{tabular}
    \caption{Table of the mean and standard deviation of $f$, (see, Eq.~\eqref{eqn:f}) along with $\sigma_0$, the first estimate of the expected evidence error provided by \texttt{dynesty}, and the $p$-value obtained from applying the Shapiro-Wilk normality test \citep{shapiro1965analysis}.}
    \label{tab:evidence}
\end{table}

\end{document}